\documentclass[aps,prb,twocolumn,showpacs,preprintnumbers,superscriptaddress]{revtex4}
\pdfoutput=1
\usepackage[intlimits]{amsmath}
\usepackage{bm,amsmath,amssymb,latexsym,mathrsfs,enumerate,color}
\usepackage{soul}
\usepackage[mathcal]{euscript}
\usepackage[hypertex]{hyperref}
\usepackage{graphicx}

\renewcommand{\vec}[1]{\bm{#1}}

\begin{document}

\title{Multiple vortex-antivortex pair generation in magnetic nanodots}

\author{Yuri Gaididei}
    \affiliation{Institute for Theoretical Physics, 03143 Kiev, Ukraine}

\author{Volodymyr P. Kravchuk}
 \email[Corresponding author. Electronic address:]{vkravchuk@bitp.kiev.ua}
    \affiliation{Institute for Theoretical Physics, 03143 Kiev, Ukraine}

\author{Denis D. Sheka}
     \affiliation{Institute for Theoretical Physics, 03143 Kiev, Ukraine}
     \affiliation{National Taras Shevchenko University of Kiev, 03127 Kiev, Ukraine}

\author{Franz G.~Mertens}
    \affiliation{Physics Institute, University of Bayreuth, 95440 Bayreuth, Germany}

\date{\today}

%
%

\begin{abstract}

The interaction of a magnetic vortex with a rotating magnetic field
causes the nucleation of a vortex--antivortex pair leading to a
vortex polarity switching. The key point of this process is the
creation of a dip, which can be interpreted as a nonlinear resonance
in the system of certain magnon modes with nonlinear coupling. The
usually observed single-dip structure is a particular case of a
multidip structure. The dynamics of the structure with $n$ dips is
described as the dynamics of nonlinearly coupled modes with
azimuthal numbers $m=0,\pm n,\pm 2n$. The multidip structure with
arbitrary number of vortex-antivortex pairs can be obtained in
vortex-state nanodisk using a space- and time-varying magnetic
field. A scheme of a possible experimental setup for multidip
structure generation is proposed.
\end{abstract}




\maketitle

\section{Introduction}
\label{sec:intro}

A magnetization curling occurs in magnetic particles of nanoscale
due to the dipole-dipole interaction. In particular, the vortex
state is realized in a disk shaped particle, where the magnetization
becomes circular lying in the disk plane in the main part of the
sample, which possesses a flux-closure state. At the disk center
there appears an out-of-plane magnetization structure (the vortex
core, typically from 10 nm \cite{Wachowiak02} to 23 nm
\cite{Chou07}) due to the dominant role of the exchange interaction
inside the core. \cite{Hubert98,Skomski03} The vortex state of
magnetic nanodots has drawn much attention because it could be used
for high-density magnetic storage and miniature sensors.
\cite{Cowburn02,Skomski03} Apart from that, such nanodots are very
attractive objects for experimental investigation of the vortex
dynamics on a nanoscale.

An experimental discovery of a vortex core reversal process by
excitation with short bursts of an alternating field
\cite{Waeyenberge06} initiated a number of studies of the core
switching process. The mechanism of the vortex switching is of
general nature; it is essentially the same in all systems where the
switching was observed.
\cite{Waeyenberge06,Xiao06,Hertel07,Lee07a,Yamada07,Kravchuk07c,Sheka07b,Liu07b,Vansteenkiste09,Weigand09}
There are two main stages of the switching process: (i) At the first
stage of the process the vortex structure is excited by the pumping,
leading to the creation of an out--of--plane dip with opposite sign
nearby the vortex. The appearance of such a deformation is confirmed
experimentally. \cite{Vansteenkiste09,Weigand09} (ii) At the second
stage, when the dip amplitude reaches the maximum possible value,
there appears a vortex--antivortex pair from the dip structure. The
further dynamics is accompanied by the annihilation of the original
vortex with the new antivortex, leading effectively to the switching
of the vortex polarity; the dynamics of this three body problem can
be described analytically. \cite{Gaididei08b,Gaididei10} While the
second stage is well understood, the physical picture of the first
stage of the switching process, which is a dip creation, is still
not clear.

The aim of the current study is to develop a theory of the dip
creation, which is a key moment in the vortex switching process. The
dip always appears as a nonlinear regime of one of the magnon modes.
In most studies the dip appears by exciting a low frequency
gyromode, which corresponds to the azimuthal quantum number $m=-1$;
the frequency of this mode $\omega_G$ lies in the sub GHz range.
Excitation of the gyromode always leads to a macrospopic motion of
the vortex as a whole; moreover the vortex has to reach some
critical velocity $v_{\text{cri}}$ of about 300 m/s in order to
switch its polarity \cite{Lee08c,Vansteenkiste08a}.

Recently we have reported about vortex core switching under the
action of a homogeneous rotating magnetic field $\vec{B}=B_x+iB_y =
B_0\exp(i\omega t)$. \cite{Kravchuk07c} The theory of the dip
creation was constructed very recently in our previous paper
\cite{Kravchuk09}, where we found a dip as a nonlinear regime of the
high frequency mode with $m=1$; the switching process for that case
is not accompanied by a vortex motion at all. In this paper we show
that the dip can be excited for any azimuthal mode $m$ by using a
\emph{nonhomogeneous} rotating magnetic field of the form
\eqref{eq:B-field}.

The paper is organized as follows. In Sec.~\ref{sec:model} we
formulate the model and describe the approach of a rotating
reference frame. Two kinds of numerical simulations are presented in
Sec.~\ref{sec:numerics}: micromagnetic \textsf{OOMMF} simulations
(Sec.~\ref{sec:OOMMF}) and spin--lattice \textsf{SLASI} simulations
in the rotating frame (Sec.~\ref{sec:SLASI}). An analytical approach
is presented in Sec.~\ref{sec:analytics}. We discuss our results in
Sec.~\ref{sec:conclusion}. In Appendix~\ref{sec:dipolar} we prove
the conservation law for the total momentum $J_z$ under the action
of the magnetostatic interaction. A possible experimental setup for
multidip structure generation is discussed in
Appendix~\ref{sec:App_field}.

\section{Model and continuum description}
\label{sec:model}

The continuum dynamics of the the spin system can be described in
terms of the magnetization unit vector $\vec{m}=\vec{M}/M_S =
\left(\sin\vartheta\cos\varphi,\sin\vartheta\sin\varphi,\cos\vartheta\right)$,
where $\vartheta$ and $\varphi$ are functions of the coordinates and
the time, and $M_S$ is the saturation magnetization. In the
subsequent text only disk shaped samples are discussed. Therefore it
is convenient to introduce dimensionless coordinates
$\vec\rho=(x,y)/L$ within the disk plane and $\zeta=z/h$ along the
disk axis, where $L$ is the disk radius and $h$ is its thickness.
The magnetization is assumed to be uniform along the $z$-axis and so
the corresponding angular coordinates of the magnetization read
$\theta(\vec\rho)=\vartheta(L\vec\rho)$ and
$\phi(\vec\rho)=\varphi(L\vec\rho)$. The energy functional of the
system under consideration consists of three terms,
\begin{equation} \label{eq:Enegry}
E = 4\pi M_S^2V\left( \mathscr{E}^{\text{ex}} +
\mathscr{E}^{\text{ms}} + \mathscr{E}^{\text{f}}\right).
\end{equation}
Here $V$ is volume of the sample. The dimensionless energy terms are
the following:
\begin{equation} \label{eq:E-ex}
\mathscr{E}^{\text{ex}} = \frac{1}{2\pi}\frac{\ell^2}{L^2}
\int\!\mathrm{d}^2\vec{\rho} \left[\left(\nabla\theta\right)^2 +
\sin^2\theta \left(\nabla\phi\right)^2\right]
\end{equation}
is the exchange energy with $\ell=\sqrt{A/4\pi M_S^2}$ being the
exchange length, $A$ being the exchange constant. The magnetostatic
energy $\mathscr{E}^{\text{ms}}$ comes from the dipolar interaction,
see Appendix \ref{sec:dipolar}, and in the continuum limit it can be
presented as a sum of three terms:
$\mathscr{E}^{\text{ms}}=\mathscr{E}_{\text{v}}^{\text{ms}}+\mathscr{E}_{\text{s}}^{\text{ms}}+\mathscr{E}_{\text{e}}^{\text{ms}}$.
Here
\begin{equation}\label{eq:en-ms-v}
\mathscr{E}_{\text{v}}^{\text{ms}}=\frac{\varepsilon}{4\pi^2}\int_\mathcal{V}\!\mathrm{d}\mathcal{V}\int_{\mathcal{V}'}\!\mathrm{d}\mathcal{V}'
\frac{\nabla\!\cdot\!\vec m(\vec\rho)\nabla\!\cdot\!\vec
m(\vec\rho')}{\sqrt{(\vec\rho-\vec\rho')^2+(\zeta-\zeta')^2}}
\end{equation}
is the energy of the interactions of volume magnetostatic charges,
where $\varepsilon=h/L$ is the disk aspect ratio and
$\int_\mathcal{V}\mathrm{d}\mathcal{V}=\int_0^1\mathrm{d}\zeta\int_0^1\mathrm{d}\rho\rho\int_0^{2\pi}\mathrm{d}\chi$
with $(\rho,\chi)$ being polar coordinates within the disk plane.
\begin{equation}\label{eq:en-ms-s}
\begin{split}
\mathscr{E}_{\text{s}}^{\text{ms}}=\frac{1}{2\pi^2\varepsilon}\int_\mathcal{S}\!\mathrm{d}\mathcal{S}\int_{\mathcal{S}'}\!&\mathrm{d}\mathcal{S}'\cos\theta(\vec\rho)\cos\theta(\vec\rho')\times\\
&\left[\frac{1}{|\vec\rho-\vec\rho'|}-\frac{1}{\sqrt{(\vec\rho-\vec\rho')^2+\varepsilon^2}}\right]
\end{split}
\end{equation}
is the energy of the interactions of charges on the upper and bottom
surfaces, where
$\int_\mathcal{S}\mathrm{d}\mathcal{S}=\int_0^1\mathrm{d}\rho\rho\int_0^{2\pi}\mathrm{d}\chi$.
\begin{equation}\label{eq:en-ms-e}
\mathscr{E}_{\text{e}}^{\text{ms}}=-\frac{\varepsilon}{2\pi^2}\int_\mathcal{V}\!\mathrm{d}\mathcal{V}\int_{\Sigma'}\!\mathrm{d}\Sigma'
\frac{\nabla\!\cdot\!\vec
m(\vec\rho)\cos[\phi(\vec\rho')-\chi']}{\sqrt{(\vec\rho-\vec\rho')^2+(\zeta-\zeta')^2}}
\end{equation}
is the energy of the interactions of edge surface charges with the
volume charges. Here $\Sigma$ is the disk edge surface and
$\int_\Sigma\mathrm{d}\Sigma=\int_0^1\mathrm{d}\zeta\int_0^{2\pi}\mathrm{d}\chi$.
Due to magnetization uniformity along the $z$-axis the distribution
of upper and bottom surface charges is antisymmetrical and the
distributions of volume and edge surface charges are uniform along
the $z$-axis. As a result the interaction energy of upper and bottom
surface charges with volume charges as well as with edge surface
charges is equal to zero. The last term in \eqref{eq:Enegry}
describes an interaction with a nonhomogeneous rotating magnetic
field, see below.

The evolution of magnetization can be described by the Landau--Lifshitz--Gilbert (LLG) equation:
\begin{subequations} \label{eq:LLG}
\begin{align} \label{eq:LLG-angular-1}
-\sin\theta\ \dot\phi &= -\pi\frac{\delta \mathscr{E}}{\delta \theta} - \eta \dot\theta,\\
\label{eq:LLG-angular-2} %
\sin\theta\ \dot\theta &= -\pi\frac{\delta \mathscr{E}}{\delta\phi}
- \eta \sin^2\theta \dot\phi.
\end{align}
\end{subequations}
Here and below the overdot indicates derivative with respect to the dimensionless time
\begin{equation} \label{eq:dimensionless}
\tau = \omega_0 t, \qquad \omega_0 = 4\pi\gamma M_S,
\end{equation}
where $\gamma$ is the gyromagnetic ratio, $\eta$ is the Gilbert
damping constant, and the factor $\pi$ appears due to the disk
volume normalization. These equations can be derived from the
following Lagrangian
\begin{equation} \label{eq:L}
\mathscr{L} = \frac{1}{\pi}\int\!\mathrm{d}^{2}\vec{\rho}
\left(1-\cos\theta \right)\dot\phi - \mathscr{E}
\end{equation}
and dissipation function
\begin{equation} \label{eq:F}
\mathscr{F} = \frac{\eta}{2 \pi}\int\!\mathrm{d}^{2}\vec{\rho}
\left({\dot\theta}^2+ \sin^2\theta{\dot\phi}^2 \right).
\end{equation}

Let us start with the no--driving case $\mathscr{E}^\mathrm{f}=0$.
The exchange interaction \eqref{eq:E-ex} provides the conservation
of the total magnetization $M_z$ along the cylindrical axis $z$ and
the conservation of the $z$--component of the orbital momentum $L_z$
\begin{equation} \label{eq:M&L}
M_z = \frac{1}{\pi}\int\!\mathrm{d}^{2}\vec{\rho}\cos\theta, \qquad
L_z=-\frac{1}{\pi}\int\!\mathrm{d}^{2}\vec{\rho}\cos\theta
\partial_{\chi}\phi
\end{equation}
due to the invariance under rotation about the $z$-axis in
spin-space and physical space, respectively. It is
well--known\cite{Papanicolaou91} that the magnetostatic interaction
breaks both symmetries. Nevertheless for thin cylindrical samples,
when the magnetization distribution does not depend on the thickness
coordinate, the magnetostatic energy is invariant under two
simultaneous rotations
\begin{equation} \label{eq:rotations}
\phi\rightarrow \phi + \varphi_0, \qquad \chi\rightarrow\chi + \varphi_0,
\end{equation}
leading to the conservation of the total momentum
\begin{equation} \label{eq:Jz}
J_z = \frac{1}{\pi}\int\!\mathrm{d}^{2}\vec{\rho}\left(\cos\theta -
1\right) \left(1- \partial_{\chi}\phi\right),
\end{equation}
see the proof in Appendix \ref{sec:dipolar}.

\subsection{Planar vortex and magnon modes}
\label{sec:vortex}

The ground state of a small size nanodisk is uniform; it depends on
the particle aspect ratio $\varepsilon$: thin nanodisks are
magnetized in the plane (when $\varepsilon
<\varepsilon_c\approx1.812$) and thick ones along the axis (when
$\varepsilon >\varepsilon_c$). When the particle size exceeds some
critical value, the magnetization curling becomes energetically
preferable due to the competition between the exchange and dipolar
interaction. For a disk shape particle there appears the vortex
state.

Let us consider the static solutions of the Landau-Lifshitz equation
with in--plane magnetization ($\theta\equiv\pi/2$). The in-plane
magnetization angle $\phi$ satisfies the Laplace equation
$\nabla^2\phi=0$. Typically for the Heisenberg magnets the boundary
is free, which corresponds to the Neumann boundary conditions.
However, the dipolar interaction orients the magnetization
tangentially to the boundary in order to decrease the surface
charges. Effectively this can be described by fixed (Dirichlet)
boundary conditions. The simplest topologically nontrivial solution
of this boundary value problem is the planar vortex, situated at the
disk center
\begin{equation} \label{eq:vortex}
\phi^{\text{v}} = \chi+\mathfrak{C}\frac{\pi}{2},
\end{equation}
where $\mathfrak{C}=\pm1$ is the vortex chirality. Such planar
vortices are known for the Heisenberg magnets with strong enough
easy--plane anisotropy \cite{Mertens00}. Qualitatively, when the
typical magnetic length is smaller than the lattice constant, the
pure planar vortex can be realized. In the magnetically soft
nanomagnets like permalloy, there exist out--of--plane vortices. The
typical out--of--plane vortex structure has a bell--shaped form with
a core size about the exchange length. \cite{Hubert98}

\begin{figure}
\includegraphics[width=0.9\columnwidth]{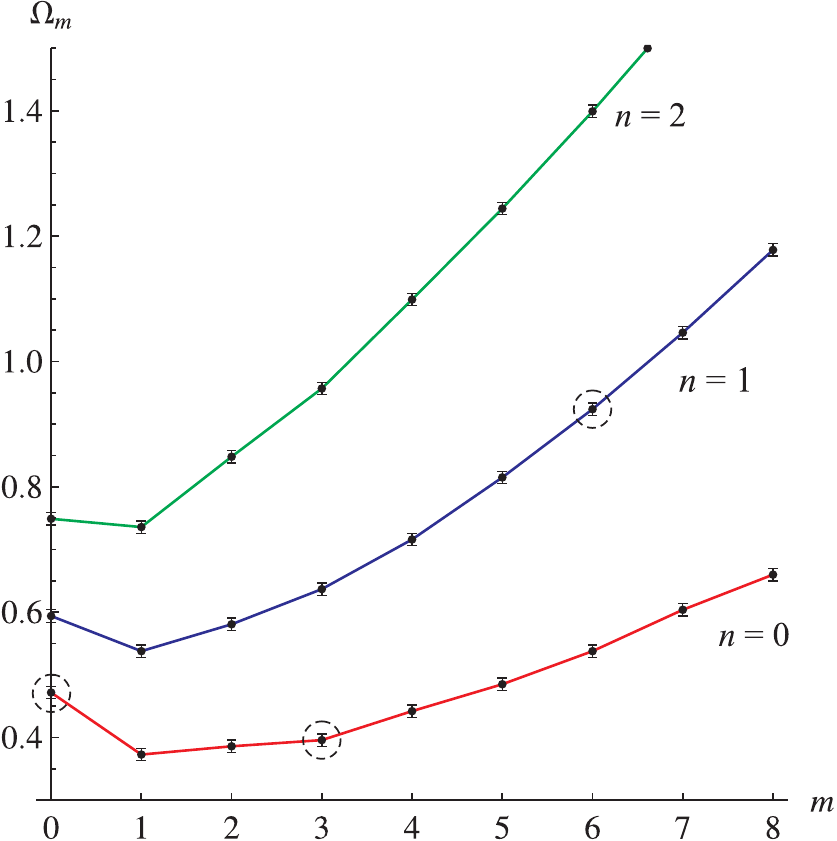}
\caption{(Color online) Eigenfrequencies of the magnon waves in a
vortex-state Py disk ($150$~nm diameter, $20$~nm thickness). The
spectrum was obtained using micromagnetic modelling with a fixed
in--plane vortex core (for technical details see
Sec.~\ref{sec:numerics}). The frequencies are normalized by
$\omega_0=4\pi\gamma M_S$ (30.3 GHz). $n$ denotes the radial wave
number -- the number of roots of the functions $f(\rho)$ and
$g(\rho)$. The separation of waves with different $m$ and $n$ was
achieved using spatio-temporal Fourier transform technique.}
\label{fig:modes} %
\end{figure}

Magnons on a vortex background can be described using the partial wave expansion:
\begin{equation} \label{eq:Ansatz}
\begin{split}
\cos\theta &=\cos\theta^{\text{v}}(\rho) + \frac{1}{\sqrt{2}}\sum_{m,n}\alpha_{m,n}(\tau)f_{|m|,n}(\rho)e^{i m \chi},\\
\phi&=\phi^{\text{v}}
+\frac{1}{\sqrt{2}}\sum_{m,n}\beta_{m,n}(\tau)g_{|m|,n}(\rho)e^{i m
\chi},
\end{split}
\end{equation}
where $\alpha_{m,n} = \alpha_{-m,n}^\star$ and $\beta_{m,n} =
\beta_{-m,n}^\star$. In the linear regime $\alpha_{m,n}(\tau) =
\exp(i\varOmega_{m,n}\tau)$ and $\beta_{m,n}(\tau) =
\exp(i\varOmega_{m,n}\tau)$. We consider here the planar vortex,
where the spectrum of eigenmodes is degenerate  with respect to the
sign of $m$. Typical eigenfrequencies are presented in
Fig.~\ref{fig:modes}. The functions $f$ and $g$ obey the following
normalization rule
\begin{equation}\label{eq:normirovka}
\langle
f_{m,n}g_{m',n'}\rangle\equiv\int_0^1f_{m,n}(\rho)g_{m',n'}(\rho)\rho\mathrm{d}\rho=\delta_{m,m'}\delta_{n,n'}.
\end{equation}
Below we assume that for each azimuthal number $m$ it is sufficient
to take into account only one radial wave with a certain radial
index $n$. Therefore the summation over $n$ will be omitted.

\begin{figure*}
\includegraphics[width=\textwidth]{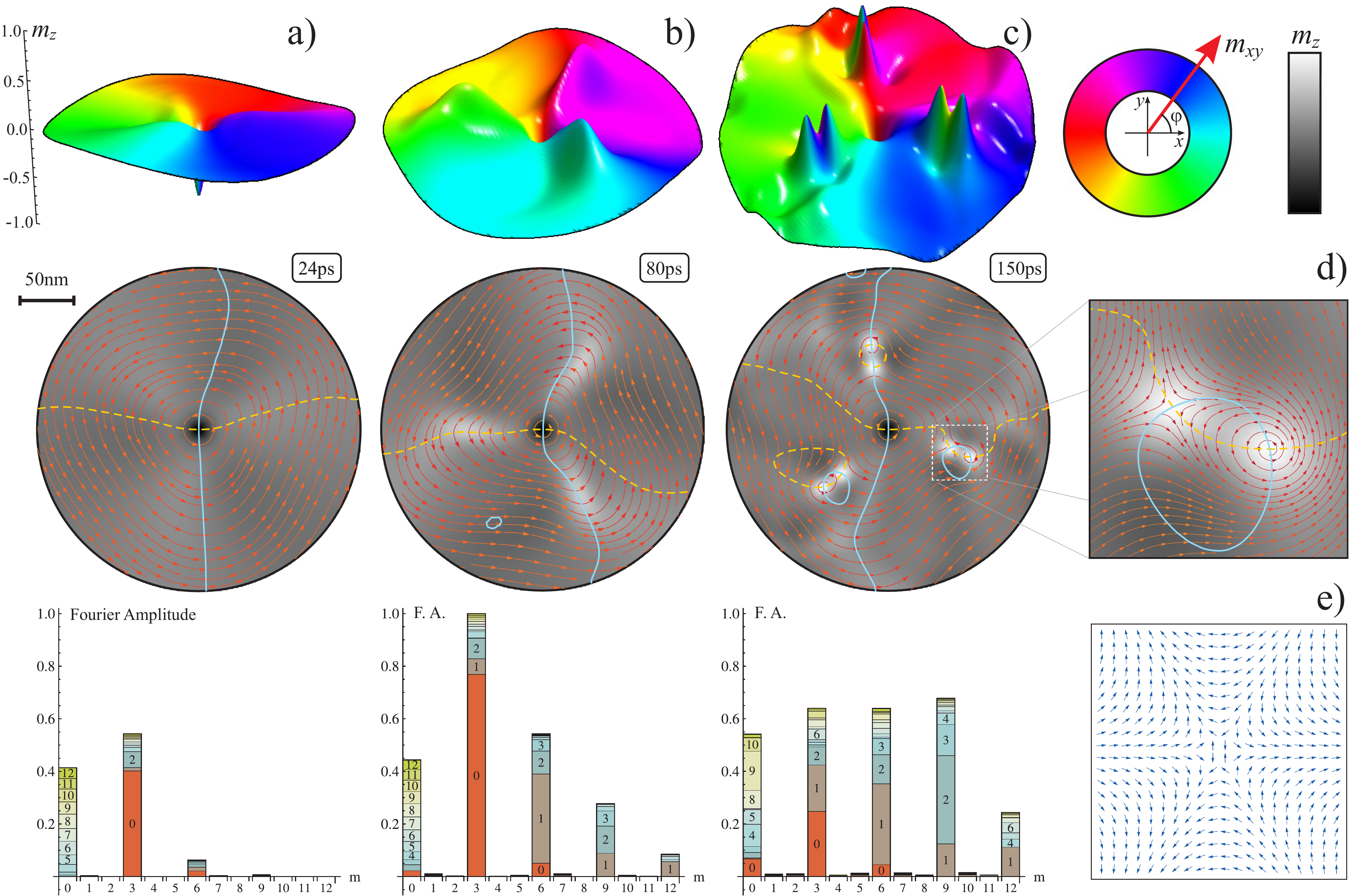}
\caption{(Color online) Vortex state dynamics under influence of the
field \eqref{eq:B-field} with $\mu=-3$, $B_0$=40mT and
$\omega$=6GHz. Disk radius $L$=150nm. Columns a), b) and c)
correspond to the different moments of time after the field
switching-on moment: 24, 80 and 150ps, respectively. The top row of
these columns illustrates distribution of the out-of-plane
magnetization component ($m_z$). The in-plane component distribution
is shown in the second row. Isolines $m_x=0$ and $m_y=0$ are shown
as dashed and solid lines, respectively. In the bottom row the
corresponding two-dimensional  spatial Fourier transforms of the
distribution $m_z(r,\chi)$ are presented. Each bar of a certain
color corresponds to a Fourier amplitude $F_{mn}$, where the
azimuthal wave number $m$ is specified by the horizontal axis and
the radial number $n$ is indicated on the bar. The subfig.~d)
demonstrates a vortex-antivortex pair in detail. The used magnetic
field at $t=0$ is shown in the subfig.~e). }
\label{fig:3-dips-dynamics}
\end{figure*}

\subsection{Interaction with a field}
\label{sec:field}

Let us consider the effects of a magnetic field. The role of the
field is to excite spin waves on a vortex background. It is well
known that the low frequency gyroscopical mode can be excited by
applying a homogeneous ac field with a frequency $\omega_G$ in the
sub GHz range. The strong pumping of such a mode is known to cause
the vortex polarity switching.

The vortex switching phenomenon was studied recently by applying a
homogeneous high frequency rotating magnetic field.\cite{Kravchuk09}
Such a field excites the azimuthal mode with azimuthal number
$m=\pm1$; the frequency $\omega_1$ of such a mode lies in the range
of 10 GHz.\cite{Kravchuk07c}

In the present work we study the vortex switching process by
exciting higher magnon modes with higher azimuthal numbers $m$. In
order to excite such a mode we propose to consider the influence of
a nonhomogeneous rotating magnetic field
\begin{equation} \label{eq:B-field}
\vec{B} = B_x + i B_y = B_0e^{i(\mu+1)\chi + i\omega t}, \qquad \mu\in\mathbb{Z}.
\end{equation}
Such a field distribution is chosen, because it directly pumps the
magnon mode with $m=\mu$. Using Ansatz \eqref{eq:Ansatz} one can
easily calculate the Zeeman energy
\begin{equation} \label{eq:Energy-f}
\begin{split}
\mathscr{E}^{\text{f}} &= -\frac{b}{\pi}\int\!\mathrm{d}^{2}\vec{\rho} \sin\theta \cos\Bigl((\mu+1)\chi + \Omega \tau - \phi\Bigr)\\
& = b_e \left(\beta_\mu e^{-i\Omega\tau}+ \beta_\mu^\star
e^{i\Omega\tau}\right), \quad b_e =
\frac{b\mathfrak{C}}{\sqrt{2}}\langle g_\mu\rangle.
\end{split}
\end{equation}
Here and below we use the normalized field intensity $b=B_0/4\pi
M_S$ and field frequency $\Omega=\omega/\omega_0$.  For the case
$\mu<-1$ the field with structure \eqref{eq:B-field} can be created
experimentally, for details see Appendix \ref{sec:App_field}.

\subsection{Rotating frame of reference}
\label{sec:rfr}

In the laboratory frame of reference the total energy $\mathscr{E}$
depends explicitly on time because of the Zeeman term
\eqref{eq:Energy-f}, which contains the explicit time dependence.
But using a transition into rotating reference frame (RRF) by the
way of
\begin{equation} \label{eq:rot-frame}
\widetilde{\phi}=\phi+\widetilde{\Omega} \tau,\qquad \widetilde{\chi}=\chi+\widetilde{\Omega}\tau, \qquad \widetilde{\Omega} = \frac{\Omega}{\mu},
\end{equation}
and using the invariance of the exchange and magnetostatic energies with respect to the simultaneous rotations \eqref{eq:rotations}, we obtain the time-independent energy in the RRF:
\begin{equation} \label{eq:E-rot}
\widetilde{\mathscr{E}} = \mathscr{E}^\mathrm{ex} + \mathscr{E}^\mathrm{ms} + \widetilde{\mathscr{E}}^\mathrm{f} - \widetilde\Omega J_z
\end{equation}
Here the Zeeman term $\widetilde{\mathscr{E}}^\mathrm{f}$ does not
contain time explicitly and has the form
\begin{equation} \label{eq:E-f-rot}
\widetilde{\mathscr{E}}^\mathrm{f}=b_e \left(\beta_\mu + \beta_\mu^\star\right).
\end{equation}
The dissipation function \eqref{eq:F} in the RRF reads
\begin{equation}
\begin{split} \label{eq:Fdis-rot}
\widetilde{\mathscr{F}}&=\mathscr{F}+\widetilde{\Omega}\mathfrak{F},\\
\mathfrak{F}&=\frac{\eta}{\pi}\int \mathrm{d}^{2}\vec\rho
\left[\dot\theta\partial_\chi\theta + \sin^2\theta(\partial_\chi\phi
- 1)\dot\phi \right].
\end{split}
\end{equation}

\section{Numerical simulations}
\label{sec:numerics}

To investigate the magnetization dynamics of a vortex state nanodisk
under influence of the external magnetic field \eqref{eq:B-field},
two kinds of simulations were used.

\subsection{Micromagnetic simulations}
\label{sec:OOMMF}

The main part of the numerical results were obtained using the full
scale \textsf{OOMMF} micromagnetic simulations with the material
parameters of Permalloy $\mathrm{Ni}_{81}\mathrm{Fe}_{19}$: exchange
constant $A=1.3\times10^{-11}\mathrm{J/m}$, saturation magnetization
$M_S=8.6\times10^5\mathrm{A/m}$, damping constant $\eta=0.01$. The
on-site anisotropy was neglected and the mesh cell was chosen to be
$2\mathrm{nm}\times2\mathrm{nm}\times h$. For all simulations the
thickness $h=20\mathrm{nm}$.

An example of magnetization dynamics induced by the field
\eqref{eq:B-field} with $\mu=-3$ is explored in
Fig.~\ref{fig:3-dips-dynamics}.

Initially we have a nanodisk in a vortex state which is a ground
state. The vortex polarity was chosen to be negative. During the
first tens of picoseconds of the field influence the generation of a
magnon mode with azimuthal number $m=-3$ is observed (see column a).
The Fourier amplitude which corresponds to $m=0$ appears due to the
out-of-plane core component of the initial vortex. Due to the
pumping the mode with $m=\mu$ goes to the nonlinear regime: areas
with $\mathrm{sign}(m_z)=\mathrm{sign}(-\mu\omega)$ become localized
which can be interpreted as dips formation (see column b). The
number of dips is equal to $|\mu|$. For small field amplitudes such
a multidip structure achieves some stationary regime and rotates
around the vortex center forever. This phenomenon is described in
detail below. Fig.~\ref{fig:3-dips-dynamics} demonstrates the
example of the magnetization dynamics for larger field amplitudes:
the vortex-antivortex pairs are nucleated from the dips (see column
c) and subfig.~d). The subsequent dynamics of the vortex-antivortex
pairs is rather complicated: the trajectory of a pair motion is not
circular, pairs radiate magnons during the motion and
self-annihilate. Then the process repeats periodically. The question
about vortex-antivortex pair motion in the presence of an immobile
central vortex is an open problem. But in this paper we focus our
attention on the dips creation mechanism only. It should be also
emphasized that the initial vortex in Fig.~\ref{fig:3-dips-dynamics}
is not pinned but it remains immobile during the dynamics. Using the
simulations we found that such a vortex stability is observed only
when the number of dips is greater than 2. The theoretical
explanation of this phenomenon is an open problem.

\begin{figure}
\includegraphics[width=\columnwidth]{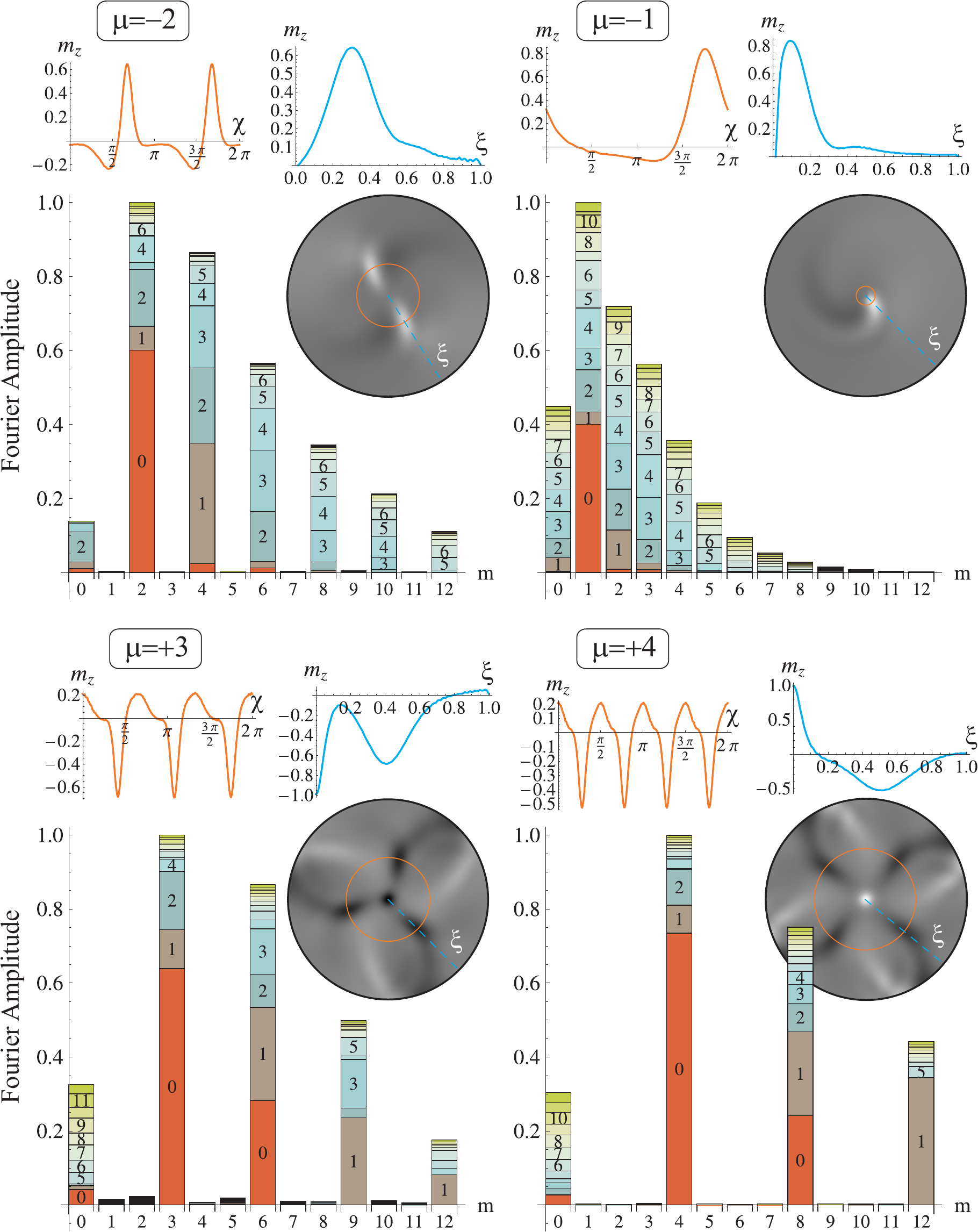}
\caption{(Color online) Multidip structures created by the field
\eqref{eq:B-field} with different $\mu$. To demonstrate the
localized character of the dips the distribution of the out-of-plane
component $m_z$ was obtained along the circle and the line $\xi$
which pass through the extremum points of a multidip structure. The
distributions along the circle and the line $\xi$ are shown above
Fourier spectra on the left and right plots, respectively. The other
notation is the same as in Fig.~\ref{fig:3-dips-dynamics}. For the
cases $\mu=-2$ and $\mu=-1$ the central vortex core is planar and
artificially fixed, for the cases $\mu=+3$ and $\mu=+4$ the central
vortex has opposite polarity and is not fixed.} \label{fig:diff_mu}
\end{figure}

Using external fields of the form \eqref{eq:B-field} with different
$\mu$ one can obtain a multidip structure with an arbitrary number
of dips. This possibility is demonstrated in Fig.~\ref{fig:diff_mu}.

It is important to emphasize the following properties of the
multidip structure formation: (i) All dips have the same polarity
which is determined by the sign of $m$ only and the dip polarity
does not depend neither on the polarity of the central vortex (see
the insets for $\mu=+3$ and $\mu=+4$) nor on the existence of an
out-of-plane vortex core at all (see the insets for $\mu=-2$ and
$\mu=-1$). (ii) the multidip structure with $n$ dips is composed by
modes with azimuthal numbers $m=n,\,2n,\,3n,\dots$ The contribution
of a mode decreases when its azimuthal number $m$ increases. (iii)
Contribution of the mode $m=0$ is essential for a single-dip
structure only (compare insets for $\mu=-2$ and $\mu=-1$).

\begin{figure}
\includegraphics[width=\columnwidth]{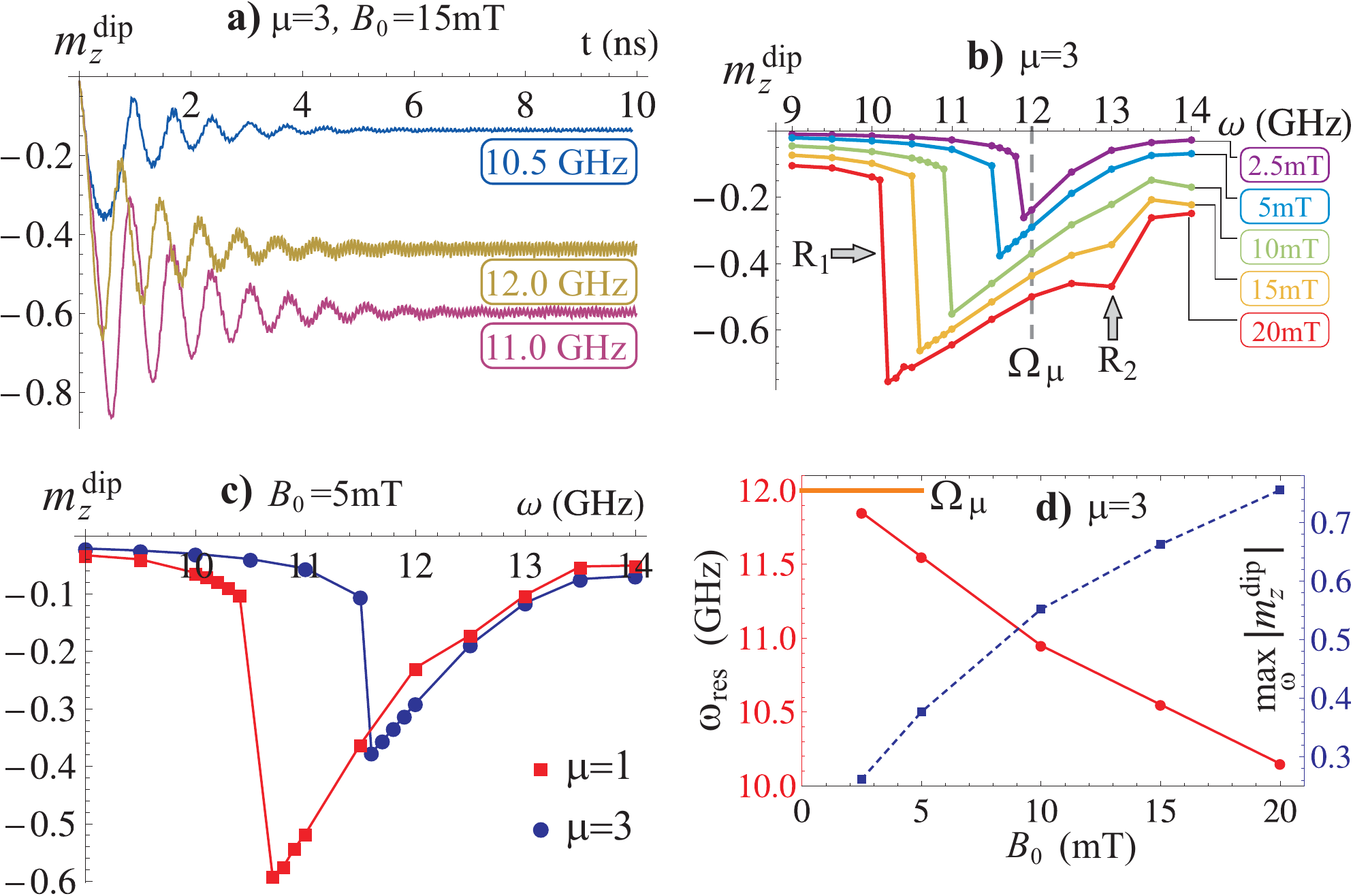}
\caption{(Color online) Characteristics of the process of multidip
structure formation. The data was obtained from simulations with a
Py disk with radius $L$=76nm and thickness $h$=20nm. If the multidip
structure is formed, then $m_z^{\mathrm{dip}}$ denotes the depth of
a dip, otherwise it denotes the amplitude of the corresponding
azimuthal mode. a) -- the process of steady-state three-dips
structure formation for a certain field strength and for different
frequencies. b) -- resonant-type dependencies of the steady-state
dips depth on the field frequency for different field amplitudes and
for a certain $\mu$. c) -- the same as b), but for a certain field
amplitude and different $\mu$. d) -- data extracted from the
resonant curves b): resonant frequency (left axis, solid line) and
maximal dips depth (right axis, dashed line) \emph{vs} applied field
amplitude.} \label{fig:res_prop}
\end{figure}

If the strength of the applied field is sufficiently small to
prevent vortex-antivortex pair formation, the magnetization dynamics
reaches some steady-state regime. This is demonstrated in
Fig.~\ref{fig:res_prop}~a) which is based on simulations with a
$\mu=+3$ field. Such a field creates potentially a structure with
three dips of negative polarity. The time dependence of the minimal
value of the $m_z$ component of the azimuthal mode with $m=\mu$ is
shown. In case of multidip structures the plotted quantity is the
depth of the dips, therefore the notation $m_z^\mathrm{dip}$ will be
used hereafter. One can see that the value $m_z^\mathrm{dip}$
reaches some steady-state level which depends on the frequency of
the applied field. This dependence has an unusual resonance
character, see Fig.~\ref{fig:res_prop}~b).

The transition from the regime of linear modes to the multidip
regime occurs sharply when the frequency of the applied field
reaches some value $\omega_\mathrm{res}$. Vertical steps on the
dependencies $m_z^\mathrm{dip}(\omega)$ in
Fig.~\ref{fig:res_prop}~b) correspond to the above indicated
transition. It should be noted that the critical frequency depends
on the field amplitude and is always smaller than the eigenfrequency
$\omega_\mu$ of the corresponding mode, see
Fig.~\ref{fig:res_prop}~d). This points to an inherently nonlinear
nature of the resonance which is discussed in Section
\ref{sec:analytics}. Similar resonances are a feature of the
multidip structures with different number of dips,
Fig.~\ref{fig:res_prop}~c).

\subsection{Spin--lattice simulations}
\label{sec:SLASI}

We have described above the micromagnetic study of the nonlinear
dynamics of the magnetization under the influence of a
time--dependent magnetic field. The main issue of this study is the
creation of a multidip structure, i.e. a stable nonlinear state of
the system, which rotates due to the field rotation. Since the total
energy of the system becomes time independent in the RRF, see
\eqref{eq:E-rot}, one can suppose that a multidip structure forms a
stationary state of the system in the RRF.

In order to check the RRF approach, we used another kind of
simulations. Namely, we performed \textsf{SLASI} simulations, an
in--house--developed spin--lattice code. \cite{Caputo07b}
\textsf{SLASI} simulations are based on the numerical solution of
the discrete version of the LLG equations \eqref{eq:LLG}
\begin{equation} \label{eq:LLG-discrete}
\begin{split}
\frac{\mathrm{d} \vec{S}_{\vec{n}}}{\mathrm{d}t}= -
\left[\vec{S}_{\vec{n}} \times \frac{\partial \mathcal{H}}{\partial
\vec{S}_{\vec{n}}}\right] - \frac{\eta}{S} \left[
\vec{S}_{\vec{n}}\times \frac{\mathrm{d} \vec{S}_{\vec{n}}
}{\mathrm{d}t}\right],
\end{split}
\end{equation}
where the 3D spin distribution is supposed to be independent of the
z coordinate. We consider Eqs.~\eqref{eq:LLG-discrete} on 2D square
lattices of size $(2L)^2$; the lattice is bounded by a circle of
radius $L$ on which the spins are free. The Hamiltonian $\mathcal{H}
= \mathcal{H}^{\text{ex}} + \mathcal{H}^{\text{dip}} +
\mathcal{H}^{\text{f}}$ is given by the Heisenberg exchange
Hamiltonian \eqref{eq:H-ex}, the dipolar energy
\eqref{eq:H-dip-via-ABC}, and the field interaction energy
$\mathcal{H}^{\text{f}}$, which is the discrete version of
\eqref{eq:Energy-f}. The 4th--order Runge--Kutta scheme with time
step $0.01/N_z$ was used for the numerical integration of
Eqs.~\eqref{eq:LLG-discrete}.

\begin{figure}
\includegraphics[width=\columnwidth]{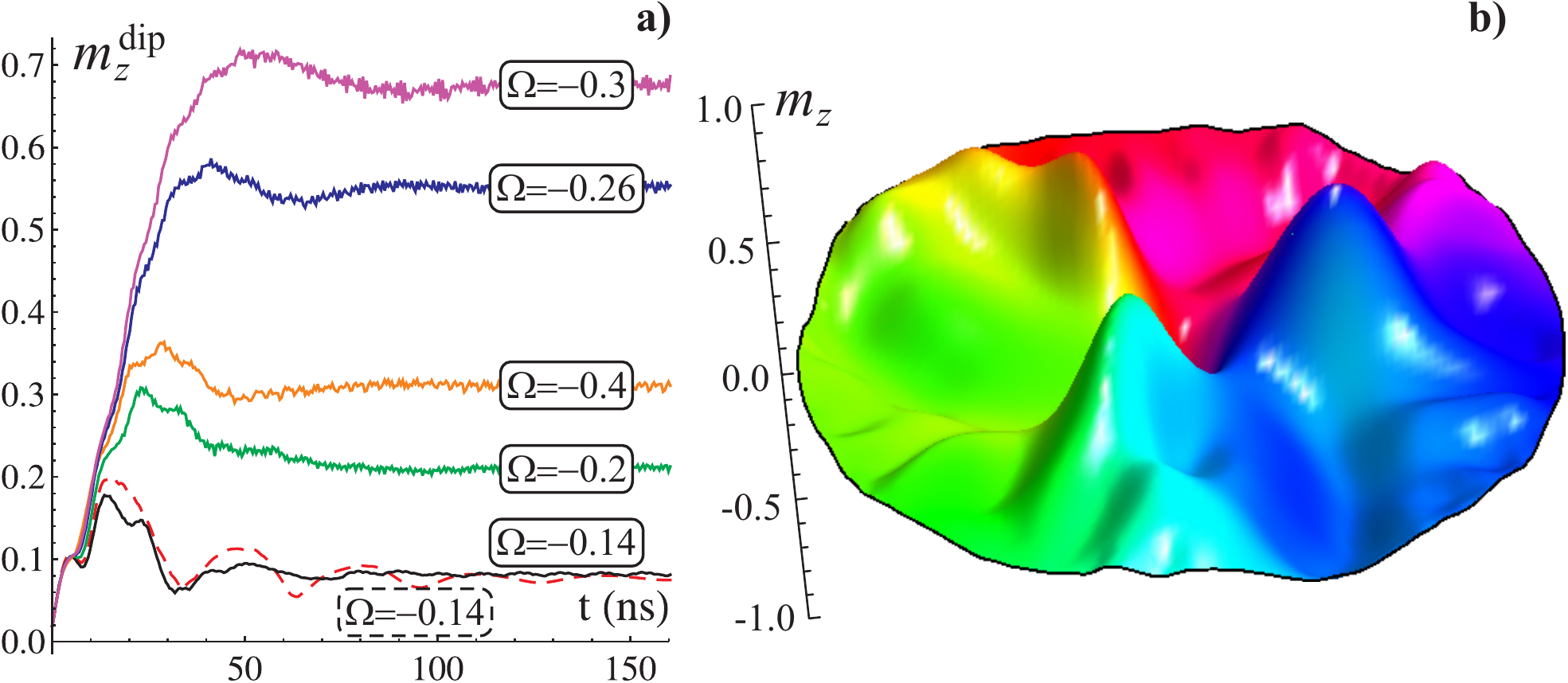}
\caption{(Color online) Multidip structure formation under the
influence of the field \eqref{eq:B-field} with $\mu=-3$ and
$b$=0.008. The data were obtained from \textsf{SLASI} simulations
for a disk with radius $L=50a$, thickness $h=5a$, exchange length
$\ell=a$, and damping coefficient $\eta=0.1$. a) Solid curves
correspond to the simulations in the laboratory reference frame for
the rotating magnetic field with different frequencies $\Omega$. The
dashed curve corresponds to the simulations in the RRF with
$\Omega^{\text{rot}} = \Omega$ under the influence of a static
field. b) Out-of-plane spin distributions from simulations in the
RRF with $\Omega^{\text{rot}} = 0.3$ at the moment $\tau \approx 24$
(same color code as in Fig.~\ref{fig:3-dips-dynamics}).}
\label{fig:slasi}
\end{figure}

First of all we performed \textsf{SLASI} simulations in the
laboratory frame of reference. Here the spin--lattice simulations
agree with our micromagnetic results. In particular, the resonance
behavior of the multidip structure is well--pronounced in
Fig.~\ref{fig:slasi}a) (solid curves), with a maximum dip amplitude
about the frequency $\Omega=0.3$.

The total energy of the magnet in the continuum limit in the RRF
\eqref{eq:E-rot} has two main differences in comparison with the
laboratory reference frame. First of all, instead of the time
dependent Zeeman energy \eqref{eq:Energy-f}, one has the influence
of a constant field with the same intensity. Apart of this, there
appears an additional rotation energy $\mathscr{E}^{\text{rot}} =
\widetilde{\Omega} J_z$. For the discrete system one can perform a
similar transformation, using the discrete Zeeman energy of the
interaction with a constant field. The discrete analogue of the
rotation energy is determined by the discrete version of the $J_z$
momentum \eqref{eq:Jz}. Results of numerical simulations in the
rotating frame of reference are plotted by the dashed curve in
Fig.~\ref{fig:slasi}a) for the rotational frequency
$\Omega^{\text{rot}} = 0.3$; they are in a good agreement with
simulations in the laboratory reference frame for $\Omega = 0.3$.
Note that we presented on Fig.~\ref{fig:slasi}a) simulations in the
RRF for small enough rotational frequencies. The reason is that
$J_z$ is not a good characteristics of the discrete system: both
exchange and dipolar interaction break the conservation of $J_z$,
since there is no rotation invariance of the discrete system.
Therefore simulations in the RRF works well when the contribution of
the $J_z$ term in the discrete Hamiltonian is small enough, and the
discreteness effects are small. This is valid for small enough
frequencies of the rotations, see Fig.~\ref{fig:slasi}a), or for
high frequencies but small enough times, see Fig.~\ref{fig:slasi}b).

\section{Analytical description of the dip creation}
\label{sec:analytics}

In order to describe the problem analytically, we consider the
problem in the RRF. As we have seen using \textsf{SLASI}
simulations, the dip can be considered as a stationary state of the
system in the RRF.

To gain some insight how the interaction with the magnetic field
(which is static in the rotating frame) together with the rotation
provides the multidip creation we use the Ansatz \eqref{eq:Ansatz}.
In order to simplify the model we consider the dip formation on a
background of a pure in--plane vortex with
$\cos\theta^{\text{v}}$=0. This approximation is confirmed by our
numerical simulations, as well as by a previous study.
\cite{Kravchuk09} The field pumps directly only the mode with
azimuthal number $m=\mu$. This mode is coupled, first of all, to the
mode with $m=-\mu,0,\pm2\mu$. Thus we will consider modes with
$m=0,\pm\mu,\pm2\mu$. Taking into account only cubic nonlinear terms
in the magnetic energy, one can use the Ansatz \eqref{eq:Ansatz} to
calculate the effective Lagrangian
\begin{equation}\label{eq:eff-L}
\mathscr{L}_{\text{eff}} = - \sum_{m=0;\pm\mu;\pm2\mu} \alpha_m^\star \dot{\beta}_m - \mathscr{E}_{\text{eff}}
\end{equation}
and the effective dissipation function
\begin{equation}\label{eq:F-eff}
\begin{split}
\mathscr{F}^{\text{eff}}=\frac{\eta}{2}\sum_m\Bigl[&\mathcal{A}_m \dot{\alpha}_m \left(\dot{\alpha}_m^\star - \frac{2im\Omega}{\mu} \alpha_m^\star\right)\\
+ &\mathcal{B}_m \dot{\beta}_m \left(\dot{\beta}_m^\star -
\frac{2im\Omega}{\mu} \beta_m^\star \right)\Bigr],
\end{split}
\end{equation}
where $\mathcal{A}_m = \langle f_{m}^2 \rangle$ and $\mathcal{B}_m =
\langle g_{m}^2 \rangle$.

The effective energy consists of several parts
$\mathscr{E}_{\text{eff}} = \mathscr{E}_{\text{eff}}^{\text{osc}} +
\mathscr{E}_{\text{eff}}^{\text{f}} +
\mathscr{E}_{\text{eff}}^{\text{rot}} +
\mathscr{E}_{\text{eff}}^{\text{int}}$, where
\begin{equation}\label{eq:en-osc}
\mathscr{E}_{\text{eff}}^{\text{osc}}  = \frac12 \sum_m
\varOmega_{|m|}\left(|\alpha_m|^2 + |\beta_m|^2 \right)
\end{equation}
describes the linear part of the modes oscillation, the energy of
the interaction with the field $\mathscr{E}_{\text{eff}}^{\text{f}}$
is equal to $\widetilde{\mathscr{E}}^{\text{f}}$ in
\eqref{eq:E-f-rot}, the rotation energy
\begin{equation}
\mathscr{E}_{\text{eff}}^{\text{rot}} = -i\frac{\Omega}{\mu} \sum_m m\alpha_m\beta_m^\star
\end{equation}
appears due to the transition into the noninertial frame of
reference, and the energy of the nonlinear coupling between the
modes has the form
\begin{equation}\label{eq:en-int}
\begin{split}
\mathscr{E}_{\text{eff}}^{\text{int}} =
i\sum_{m,n}
m(k_{m,n}^\alpha\alpha_m\alpha_n+k_{m,n}^\beta\beta_m\beta_n)\beta_{m+n}^\star.
\end{split}
\end{equation}
For the explicit form of the magnon frequencies $\varOmega_m$ and
the nonlinearity coefficients  $k_{i,j}^\xi$ see Appendix
\ref{sec:frecs_and_nonlin}.

\begin{figure}
\includegraphics[width=\columnwidth]{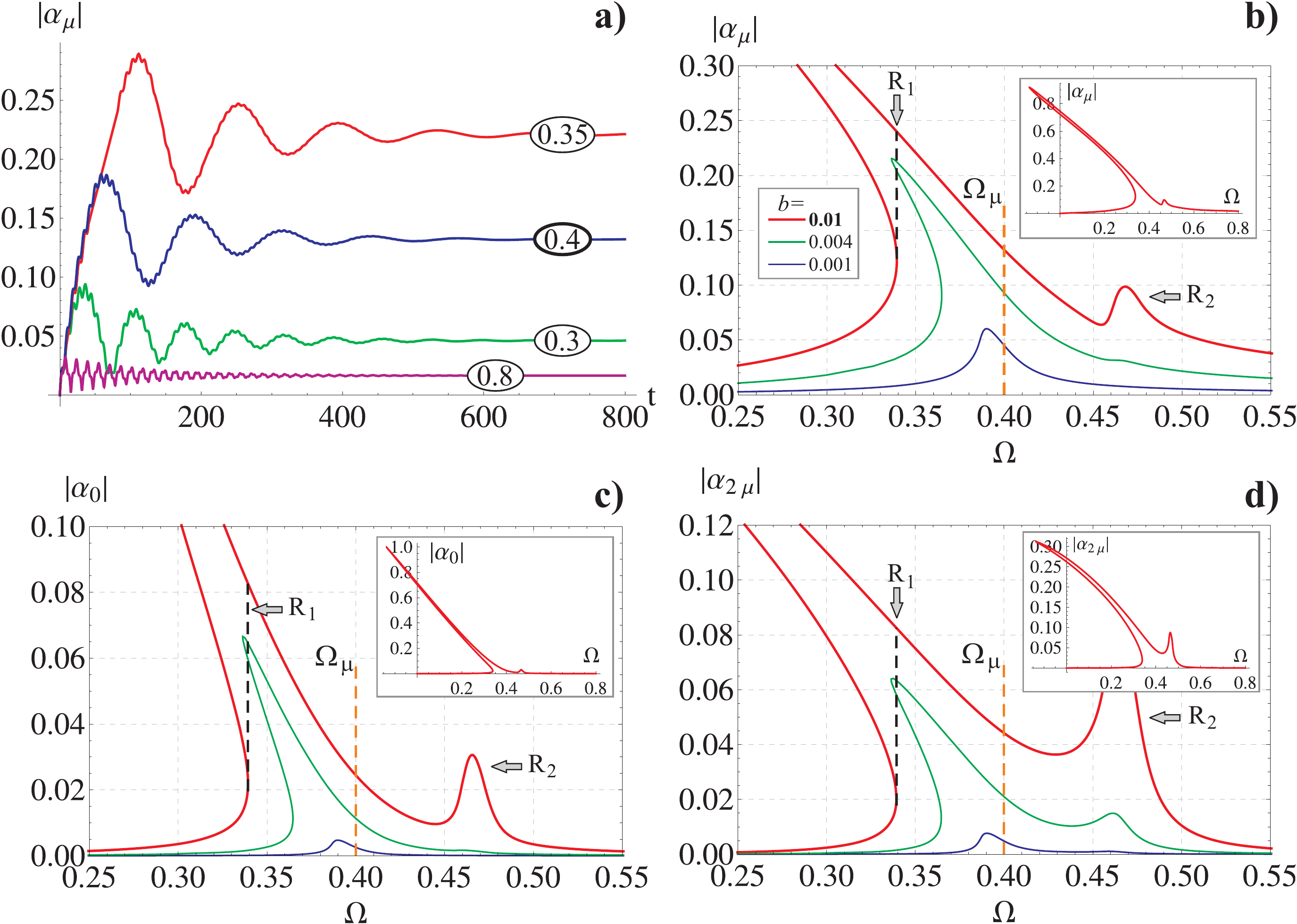}
\caption{(Color online) The nonlinear resonance in system
\eqref{eq:EOM}. Subfigure a) demonstrates time dependencies of the
amplitude of the out-of-plane component of the mode with $m=\mu$.
The shown dependencies were obtained numerically directly from
\eqref{eq:EOM} for different field frequencies $\Omega$ (the values
are in the oval frames). The values of the other parameters were the
following: $\mu=3$, $\varOmega_\mu=0.4$, $\varOmega_0=0.47$,
$\varOmega_{2\mu}=0.92$, $b=0.01$, $\eta=0.02$, $k=0.1$,
$\mathcal{A}_{i}=\mathcal{B}_{i}=1$. Amplitude-frequency
characteristics for different modes are presented in the subfigures
b)-d). The numerical values of the parameters are the same as in
subfigure a). The thick (red) curve is build for the field amplitude
$b=0.01$, thus it fully corresponds to the subfig. a). The small
insets shows the mentioned curve in full scale.}
\label{fig:nl-resonance}
\end{figure}

Assuming that the nonlinear coefficients take approximately the same
values $k_{i,j}^{\xi}\approx k$ one can obtain the equations of
motion for the amplitudes $\alpha_m$, $\beta_m$ in the form
\begin{equation} \label{eq:EOM}
\begin{split}
\dot\alpha_m=&\varOmega_{m}\beta_m+b_e(\delta_{m,\mu}+\delta_{m,-\mu})-i\Omega\frac{m}{\mu}\alpha_m\\
-&ik\sum_{m'}{m'}\alpha_{m'}^\star\alpha_{{m'}+m}+\eta\mathcal{B}_m\left(\dot\beta_m+i\frac{m}{\mu}\Omega\beta_m\right),\\
\dot\beta_m=&-\varOmega_{m}\alpha_m-i\Omega\frac{m}{\mu}\beta_m+ik\sum_{m'}({m'}+m)\alpha_{m'}^\star\beta_{{m'}+m}\\
-&\eta\mathcal{A}_m\left(\dot\alpha_m+i\frac{m}{\mu}\Omega\alpha_m\right).
\end{split}
\end{equation}
In the infinite set of equations \eqref{eq:EOM} we restrict
ourselves to the finite number of equations which do not contain any
amplitudes $\alpha_i,\beta_i$, with the exception of amplitudes with
indices $i=0,\pm\mu,\pm2\mu$. The obtained system was solved
numerically for the case $\mu=3$ and different field amplitudes. The
values of the corresponding eigenfrequencies
$\varOmega_{0,\mu,2\mu}$ were chosen to be equal to the ones
obtained from extra micromagnetic simulations of the magnon
dynamics, these values are marked by dashed circles in
Fig.~\ref{fig:modes}. The frequencies $\varOmega_0$ and
$\varOmega_{|\mu|}$ correspond to the lowest modes with radial
number $n=0$ while $\varOmega_{|2\mu|}$ corresponds to the mode with
$n=1$, because of two reasons: (i) according to the Fourier spectra
in Fig.~\ref{fig:3-dips-dynamics} and Fig.~\ref{fig:diff_mu} the
mode with this radial number dominates among modes with azimuthal
number $m=2\mu$; (ii) as it will be shown later the second resonance
$R_2$ (see Fig.~\ref{fig:res_prop},b) appears at the frequency
$\varOmega_{2\mu}/2$ which corresponds to the mode with $n=1$ (see
Fig.\ref{fig:modes}). According to the mentioned spectra the
amplitude of the dips in a multidip structure is determined mainly
by the value $\alpha_\mu$, i.e. the amplitude of the out-of-plane
component of the mode with $m=\mu$. The numerically obtained
dependencies $|\alpha_\mu(t)|$ are shown in the
Fig.~\ref{fig:nl-resonance}a). These dependencies are in good
agreement with the behavior of the dips depth, obtained using
simulations, see Fig.~\ref{fig:res_prop}a). Moreover, the
steady-state value of $\alpha_\mu$ has the same properties as the
steady-state dips depth $m_z^\mathrm{dip}$: (i) its frequency
dependence has a resonance character, (ii) the resonance frequency
is lower than the eigenfrequency, see Fig.~\ref{fig:nl-resonance}a).
Amplitude-frequency characteristics obtained numerically from model
\eqref{eq:EOM} (Fig.~\ref{fig:nl-resonance}b-d) let us conclude that
the phenomenon of an abrupt appearance of a dip (multidip) structure
in a vortex state is a nonlinear resonance in a system of
nonlinearly coupled modes with $m=0,\pm\mu,\pm2\mu$. The
corresponding resonance transition is shown as $R_1$ in
Fig.~\ref{fig:res_prop}b) and in Fig.~\ref{fig:nl-resonance}b)-d).

In the no--damping and weakly nonlinear limit ($k\ll1$) the
stationary values $\bar{\alpha}_m$ and $\bar{\beta}_m$ can be
represented approximately as follows:
\begin{equation} \label{eq:alpha-stat&beta-stat}
\begin{split}
\bar{\alpha}_\mu &= -\frac{ib_e\Omega}{\Omega^2-\varOmega_\mu^2}, \qquad \bar{\beta}_\mu = i\bar{\alpha}_\mu \frac{\varOmega_\mu}{\Omega},\\
\bar{\alpha}_0 &= - \frac{2k \mu\varOmega_\mu \left|\bar{\alpha}_\mu\right|^2}{\Omega\varOmega_0} = \frac{-2k\mu b_e^2 \varOmega_\mu \Omega}{\varOmega_0 \left(\Omega^2-\varOmega_\mu^2\right)^2}, \quad \bar{\beta}_0 = 0,\\
\bar{\alpha}_{2\mu} &= \frac{k\mu \bar{\alpha}_\mu^2}{\Omega} \frac{2\Omega^2+ \varOmega_\mu \varOmega_{2\mu}}{4\Omega^2-\varOmega_{2\mu}^2},\quad
\bar{\beta}_{2\mu} = ik\mu \bar{\alpha}_\mu^2 \frac{2\varOmega_\mu + \varOmega_{2\mu}}{4 \Omega^2-\varOmega_{2\mu}^2}.
\end{split}
\end{equation}
According to the last two equations an additional resonance is
expected for the frequency $\Omega=\varOmega_{2\mu}/2$. This
resonance is observed in the simulation results (see
Fig.~\ref{fig:res_prop}, b) and it is well recognized in
Fig.~\ref{fig:nl-resonance}~b-e, is denoted as $R_2$.

\section{Conclusions}
\label{sec:conclusion}

We have presented a detailed study of the dip structure generation,
which always precedes the vortex polarity switching phenomenon. The
physical reason for the dip creation is softening of a magnon mode
and consequently a nonlinear resonance in the system of certain
magnon modes with nonlinear coupling. The usually observed
single-dip structure is a particular case of a multidip structure.
The dynamics of the structure with $n$ dips can be strictly
described as the dynamics of nonlinearly coupled modes with
azimuthal numbers $m=0,\pm n,\pm 2n$. The multidip structure with an
arbitrary number of dips (or vortex-antivortex pairs) can be
obtained in a vortex-state nanodisk using a space- and time-varying
magnetic field of the form \eqref{eq:B-field}. A scheme of a
possible experimental setup for multidip structure generation is
proposed in Appendix~\ref{sec:App_field}.

\acknowledgments

The authors thank H.~Stoll for helpful discussions. The authors
acknowledge support from Deutsches Zentrum f{\"u}r Luft- und
Raumfart e.V., Internationales B{\"u}ro des BMBF in the frame of a
bilateral scientific cooperation between Ukraine and Germany,
project No.~UKR~08/001. Yu.G., V.P.K. and D.D.S. thank the
University of Bayreuth, where a part of this work was performed, for
kind hospitality. D.D.S. acknowledges support from the grant
No.~F25.2/081 from the Fundamental Researches State Fund of Ukraine.

\appendix

\section{Dipolar interaction and the conservation of the total momentum}
\label{sec:dipolar}

In this Appendix we consider a ferromagnetic system described by the
classical Heisenberg isotropic exchange Hamiltonian
\begin{equation} \label{eq:H-ex}
\mathcal{H}^{\text{ex}} =-\frac{J}{2} \sum_{\left(\vec{n},\vec{\delta}\right)} \vec{S}_{\vec{n}}\cdot \vec{S}_{\vec{n}+\vec{\delta}}
\end{equation}
and the dipolar interaction $\mathcal{H}^{\text{dip}}$:
\begin{equation} \label{eq:H-dip}
\mathcal{H}^{\text{dip}} = \frac{D}{2} \sum_{\substack{\vec{n}, \vec{n}'\\\vec{n}\neq \vec{n}'}} \frac{\vec{S}_{\vec{n}}\cdot\vec{S}_{\vec{n}'}- 3 \left(\vec{S}_{\vec{n}}\cdot \vec{e}_{\vec{n} \vec{n}'} \right) \left(\vec{S}_{\vec{n}'}\cdot \vec{e}_{\vec{n} \vec{n}'} \right)}
{|\vec{n} -\vec{n}'|^3}.
\end{equation}
Here $\vec{S}_{\vec{n}}\equiv\left(S^x_{\vec{n}}, S^y_{\vec{n}},
S^z_{\vec{n}}\right)$ is a classical spin vector with fixed length
$S$ in units of action on the site $\vec{n}=(n_x,n_y, n_z)$ of a
three--dimensional cubic lattice with integers $n_x$, $n_y$, $n_z$,
$J$ is the exchange integral, the parameter $D = \gamma^2/a^3$ is
the strength of the long--range dipolar interaction,
$\gamma=g|e|/(2mc)$ is the gyromagnetic ratio, $g$ is the
Land\'{e}--factor, $a$ is the lattice constant; the vector
$\vec{\delta}$ connects nearest neighbors, and $\vec{e}_{\vec{n}
\vec{n}'} \equiv \left(\vec{n} -\vec{n}'\right)/\left|\vec{n}
-\vec{n}'\right|$ is a unit vector.

Our main approximation is that $\vec{S}_{\vec{n}}$ depends only on
the $x$ and $y$ coordinates. Such a plane--parallel spin
distribution is adequate for thin films with a constant thickness
$h=N_z a$ and nanoparticles with small aspect ratio. Using the above
mentioned approximation the dipolar Hamiltonian can be written as
follows: \cite{Caputo07b}
\begin{equation} \label{eq:H-dip-via-ABC}
\begin{split}
&\mathcal{H}^{\text{dip}} = -\frac{D}{2} \sum_{\substack{{\vec{\nu}},{\vec{\nu'}}}} \Bigl[
A_{{\vec{\nu}}{\vec{\nu'}}}\left(\vec{S}_{{\vec{\nu}}}\cdot \vec{S}_{{\vec{\nu'}}} - 3 S_{{\vec{\nu}}}^zS_{{\vec{\nu'}}}^z\right)\\
&+ B_{{\vec{\nu}}{\vec{\nu'}}} \left( S_{{\vec{\nu}}}^x S_{{\vec{\nu'}}}^x - S_{{\vec{\nu}}}^y S_{{\vec{\nu'}}}^y \right) + C_{{\vec{\nu}}{\vec{\nu'}}}\left( S_{{\vec{\nu}}}^x S_{{\vec{\nu'}}}^y + S_{{\vec{\nu}}}^y S_{{\vec{\nu'}}}^x
\right)\Bigr]\!.\!\!\!
\end{split}
\end{equation}
Here the sum runs only over the 2D lattice. All the information
about the original 3D structure of our system is in the coefficients
$A_{{\vec{\nu}}{\vec{\nu'}}}$, $B_{{\vec{\nu}}{\vec{\nu'}}}$ and
$C_{{\vec{\nu}}{\vec{\nu'}}}$,
\begin{equation} \label{eq:A-B-C}
\begin{split}
A_{{\vec{\nu}}{\vec{\nu'}}}&=\frac12 \sum_{\substack{{n'}_z,n_z\\r_{\vec{n}\vec{n'}}\neq0}} \frac{r_{\vec{n}\vec{n'}}^2 - 3z_{\vec{n}\vec{n'}}^2}{r_{\vec{n}\vec{n'}}^5}, \\ %
B_{{\vec{\nu}}{\vec{\nu'}}}&= \frac32 \sum_{\substack{{n'}_z,n_z\\r_{\vec{n}\vec{n'}}\neq0}} \frac{x_{\vec{n}\vec{n'}}^2 - y_{\vec{n}\vec{n'}}^2}{r_{\vec{n}\vec{n'}}^5}, \\ %
C_{{\vec{\nu}}{\vec{\nu'}}}&= 3 \sum_{\substack{{n'}_z,n_z\\r_{\vec{n}\vec{n'}}\neq0}} \frac{x_{\vec{n}\vec{n'}} y_{\vec{n}\vec{n'}}}{r_{\vec{n}\vec{n'}}^5},
\end{split}
\end{equation}
where we used the notations: $x_{\vec n \vec {n'}} = a(n_x -
{n'}_x)$, $y_{\vec n \vec {n'}} = a (n_y - {n'}_y)$, $z_{\vec n \vec
{n'}} = a (n_z - {n'}_z)$, $\rho_{{\vec{\nu}} {\vec{\nu'}}} =
\sqrt{x_{\vec n\vec {n'}}^2 + y_{\vec n \vec {n'}}^2}, \; r_{\vec
n\vec {n'}} = \sqrt{\rho_{{\vec{\nu}} {\vec{\nu'}}}^2 + z_{\vec
n\vec {n'}}^2}$.

The continuum description of the system is based on smoothing the
lattice model, using the normalized magnetization $\vec{m}(\vec{r})
= \left({g\mu_B}/{a^3M_S}\right) \sum_{\vec{n}} \vec{S}_{\vec{n}}
\delta(\vec{r} - \vec{r}_{\vec{n}})$. Then the exchange energy
$\mathscr{E}^{\text{ex}}$ (normalized by $4\pi M_S^2$), the
continuum version of \eqref{eq:H-ex}, takes the form
\eqref{eq:E-ex}. The normalized magnetostatic energy, which is the
continuum version of \eqref{eq:H-dip-via-ABC}, is
\begin{equation} \label{eq:E-ms}
\begin{split}
&\mathscr{E}^{\text{ms}} = -\frac{1}{16\pi} \int \mathrm{d}^3\vec{r}\int \mathrm{d}^3\vec{r'} \frac{\mathcal{W}\left[\vec{m},\vec{m'}\right]}{R^5},\\
&\mathcal{W}\left[\vec{m},\vec{m'}\right] = \left( R^2-3|z-z'|^2\right) \Bigl[\sin\theta\sin\theta'\cos\left(\phi-\phi'\right)\\
& - 2 \cos\theta\cos\theta' \Bigr] + 3\sin\theta\sin\theta'\Bigl[ \rho^2\cos\left( \phi + \phi' - 2\chi\right)\\
& + {\rho'}^2\cos\left( \phi + \phi' - 2\chi'\right) - 2\rho \rho' \cos\left( \phi + \phi' - \chi - \chi' \right)\Bigr],\\
&R\left(\vec{r},\vec{r'}\right) = \sqrt{\rho^2 + {\rho'}^2 - 2\rho\rho'\cos(\chi-\chi')+ (z-z')^2}.
\end{split}
\end{equation}
The magnetostatic energy in the form \eqref{eq:E-ms} is invariant
under to simultaneous rotations of $\phi$ and $\chi$ with the same
constant angle $\varphi_0$, see Eq.~\eqref{eq:rotations}.

The consequence of such an invariance is the conservation of the
total momentum \eqref{eq:Jz}. Let us show explicitly that $J_z$ is
conserved. The time derivative of the total momentum
\begin{equation} \label{eq:dJzdt}
\begin{split}
\frac{\mathrm{d}J_z}{\mathrm{d}t}
& = \int\!\mathrm{d}^3\vec{r} \partial_\chi\left(\frac{1-\cos\theta}{\sin\theta}\cdot \frac{\partial \mathscr{E}}{\partial \theta}\right)\\
& + \int\!\mathrm{d}^3\vec{r} \left(\cos\theta-1\right) \left[\partial_\chi,\partial_t\right]\phi,\\
& + \int\!\mathrm{d}^3\vec{r}\left( \frac{\delta\mathscr{E}}{\delta \phi} - \frac{\delta\mathscr{E}}{\delta \phi}\partial_\chi\phi - \frac{\delta\mathscr{E}}{\delta \theta}\partial_\chi\theta \right)
\end{split}
\end{equation}
where we used an explicit form of Eqs.~\eqref{eq:LLG} in the case of
absence of magnetic field and damping. The first term in
\eqref{eq:dJzdt} vanishes due to the cylindrical symmetry of the
sample. The second term contains the commutator
$\left[\partial_\chi,\partial_t\right]\phi$; it can take
nonvanishing values for the singular field distributions like 2D
solitons and vortices with $\phi=q\chi+\text{const}$, which results
in $[\partial_x,\partial_y]\phi=2\pi
q\delta(\vec{r})$.\cite{Papanicolaou91}  Nevertheless this
singularity does not influence the last term in \eqref{eq:dJzdt} due
to the vanishing factor $(\cos\theta-1)$ at the singularity point.
This replacement $\cos\theta\to \cos\theta-1$ corresponds to the
regularization of the Lagrangian \cite{Sheka06e}.

Let us discuss the last term in \eqref{eq:dJzdt}. Since an isotropic
exchange interaction allows the conservation of $M_z$ and $L_z$
separately, we need to discuss here the influence of the
magnetostatic interaction only. Using the explicit form
\eqref{eq:E-ms}, one can rewrite \eqref{eq:dJzdt} as follows
\begin{equation*}
\frac{\mathrm{d}J_z}{\mathrm{d}t} = \frac{1}{8\pi} \!\!\int\!\! \frac{\mathrm{d}^3\vec{r} \mathrm{d}^3\vec{r'}}{R^5} \Biggl[ \frac{\partial \mathcal{W}} {\partial \theta} \partial_\chi\theta  + \frac{\partial \mathcal{W}} {\partial \phi} \left( \partial_\chi\phi - 1 \right) \Biggr] = \frac{I_1 + I_2}{8\pi}.
\end{equation*}
Here the contribution $I_1$ takes the form:
\begin{equation*}
I_1  = \!\int\!\! \frac{\mathrm{d}^3\vec{r} \mathrm{d}^3\vec{r'}}{R^5} \frac{\partial \mathcal{W}} {\partial \chi} \!=\! \int\!\! \mathrm{d}^3\vec{r} \mathrm{d}^3\vec{r'} \Biggl[\! \frac{\partial}{\partial \chi}\!\! \left(\frac{\mathcal{W}}{R^5}\right) - \mathcal{W} \! \frac{\partial}{\partial \chi} \!\!\left(\frac{1}{R^5}\right)\Biggr].
\end{equation*}
The first integral vanishes due to periodicity on $\chi$ and
cylindrical symmetry. The derivative in the last term
\begin{equation*}
\frac{\partial}{\partial \chi} \!\!\left(\frac{1}{R^5}\right) = -\frac{5 \rho\rho'}{R^7}\sin \left( \chi-\chi' \right)
\end{equation*}
is asymmetric with respect to the replacement $\vec{r}\leftrightarrow \vec{r'}$, hence this term vanishes after the integration, and $I_1=0$.

The integral $I_2  = \int \mathrm{d}^3\vec{r} \mathrm{d}^3\vec{r'} F[\vec{r},\vec{r'}]$ has the asymmetrical kernel
\begin{equation*}
\begin{split}
F[\vec{r},\vec{r'}] &=\sin\theta\sin\theta'\Bigl[ \left( R^2-3|z-z'|^2\right) \sin\left( \phi - \phi'\right)\\
& + {\rho'}^2 \sin \left( \phi + \phi' - 2\chi'\right) - \rho^2 \sin\left( \phi + \phi' - \chi - \chi' \right)\Bigr]\\
& = -F[\vec{r'},\vec{r}],
\end{split}
\end{equation*}
therefore, $I_2=0$.

Finally, one can state that the total momentum $J_z$ is conserved
for a cylindrical sample under the action of magnetostatic
interaction. One should note that the conservation of $J_z$ is known
for the local model of the magnetostatic interaction
$\mathscr{E}_{\text{loc}}^{\text{ms}} =\varkappa
\int\!\mathrm{d}^3\vec{r} \left(\nabla\cdot \vec{m}\right)^2$.
\cite{Papanicolaou91}

\section{Experimental possibility of a multidip structure creation}
\label{sec:App_field} %

\begin{figure}
\includegraphics[width=\columnwidth]{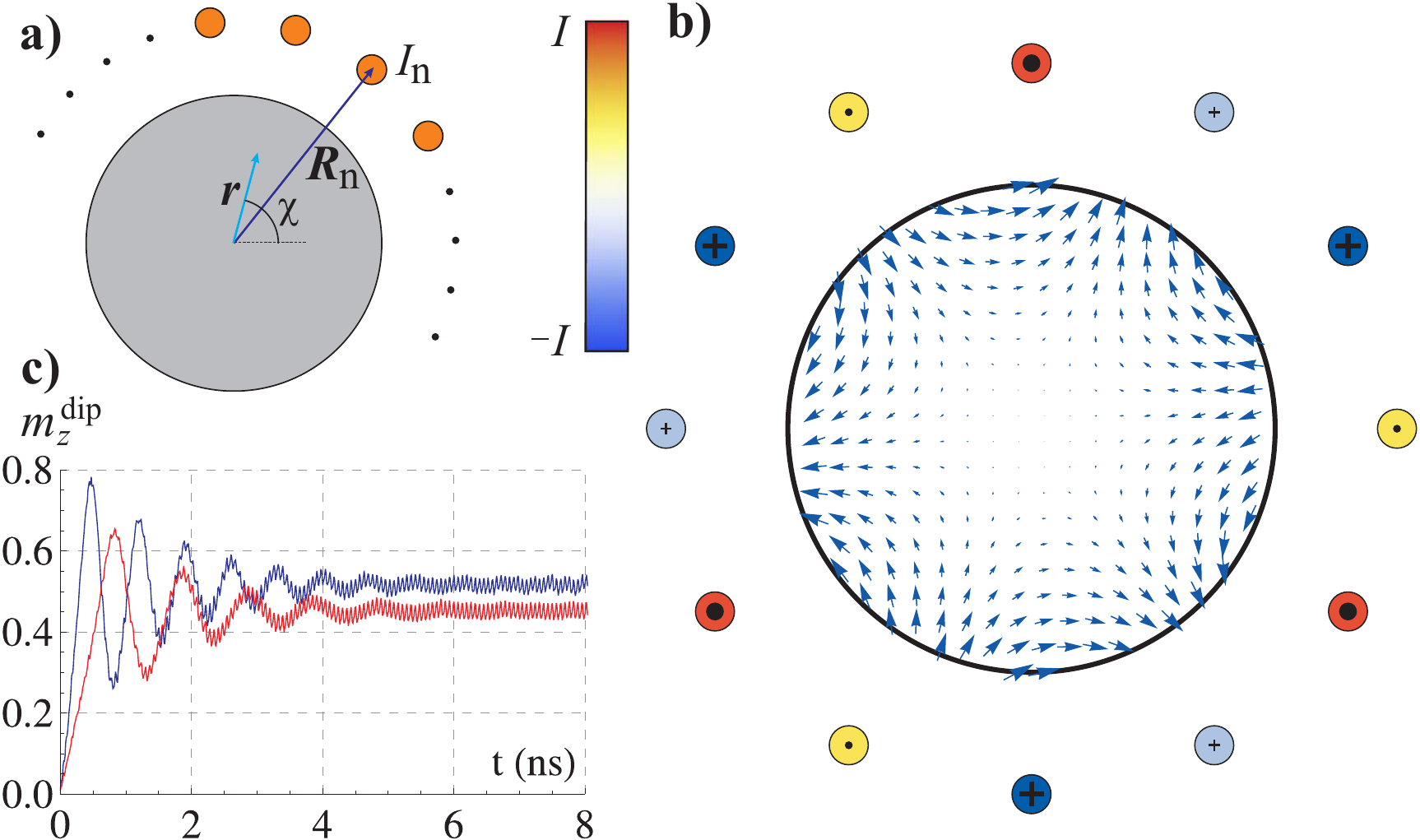}
\caption{(Color online) Possible implementation of an experimental
facility for pumping of a selected azimuthal magnon mode. Subfig. a)
demonstrates the utility assembling. Subfig.~b) shows the magnetic
field \eqref{eq:field_exact} within the disk area. The number of the
wires is $N=12$ and the phase shift of the currents is
$\Delta\phi=\pi/2$ (corresponds to $m=3$, see text). Blue and red
lines in subfig.~c) illustrate dependencies $m_z^{\mathrm{dip}}(t)$
obtained using the field \eqref{eq:B-field} with $\mu=-3$ (see
Fig.~\ref{fig:3-dips-dynamics} e) and the field \eqref{eq:Bapprox}
with $m=3$, respectively. In both cases the field amplitude was 15mT
at the disk edge. $L=76$nm, $h=$20nm.} \label{fig:field_exp}
\end{figure}

Let us consider a set of long conductive wires orientated
perpendicular to the disk plane. Let all these wires be uniformly
spaced along a circle which is concentric with the disk and has
radius a $R>L$. The described configuration is shown in
Fig.~\ref{fig:field_exp}~a).

Let the $n$-th wire conduct the current $I_n=I\cos(\omega t+2\pi m
n/N)$, where $N$ is the total number of the wires, $m$ is the
azimuthal number of the mode we aim to excite and $\omega$ should be
close to the corresponding eigenfrequency $\Omega_m$. The described
set of wires produces a magnetic field of the form
\begin{equation}\label{eq:field_exact}
\vec{B}(\vec r,\chi)=\frac{1}{c}\sum\limits_{n=0}^{N-1}I_n\frac{[\vec{e}_z\times(\vec r-\vec R_n)]}{(\vec r-\vec R_n)^2},
\end{equation}
where $\vec R_n=R[\cos(2\pi n/N),\,\sin(2\pi n/N),0]$ is the
radius-vector of the $n$-th wire, $(r,\chi,z)$ are coordinates of a
cylindrical frame of reference with $\vec e_z$ being oriented
directly to the reader, and $c$ denotes the speed of light. For the
case $N\gg m$ one can proceed from the sum in \eqref{eq:field_exact}
to an integral. And for the case $L\ll R$ the obtained integral
yields
\begin{subequations}\label{eq:Bapprox}
\begin{align}
&B_x(\xi,\chi)= \frac{NI}{2Rc}\Bigl\{\nonumber\\
&\sin[\omega t-(m-1)\chi]\left[\xi^{|m-1|}-\xi^{|m|+1}\right]-\nonumber\\
&\sin[\omega t-(m+1)\chi]\left[\xi^{|m+1|}-\xi^{|m|+1}\right]\Bigr\},\\%
&B_y(\xi,\chi)= -\frac{NI}{2Rc}\Bigl\{\nonumber\\
&\cos[\omega t-(m-1)\chi]\left[\xi^{|m-1|}-\xi^{|m|+1}\right]+\nonumber\\
&\cos[\omega t-(m+1)\chi]\left[\xi^{|m+1|}-\xi^{|m|+1}\right]\Bigr\},
\end{align}
\end{subequations}
where $\xi=r/R\ll1$. The field \eqref{eq:Bapprox} corresponds to the
field \eqref{eq:B-field} with $\mu=-|m|$ and
$\mathrm{sgn}(\omega)=\mathrm{sgn}(m)$,  but contrary to
\eqref{eq:B-field} it is radially dependent. The radial dependence
does not change the ability of the field \eqref{eq:Bapprox} for the
production of multidip structures. The radial dependent field
results only in a small decreasing of the dips amplitude, see
Fig.\ref{fig:field_exp} c). The accuracy of conformity of
\eqref{eq:Bapprox} with \eqref{eq:field_exact} strongly depends on
$N$. Using the minimal number of the wires $N=2m$ one can excite the
standing wave with azimuthal number $m$. To obtain the multidip
structure one should take $2m<N\le4m$. It is senseless to take
$N>4m$ because it does not improve the accuracy significantly.
Satisfactory accuracy was achieved for $L/R<1/2$.

\section{Microscopic expressions for magnon frequencies and nonlinearity coefficients}
\label{sec:frecs_and_nonlin} Substitution of the Ansatz
\eqref{eq:Ansatz} into \eqref{eq:E-ex} and integration results in:
\begin{equation}\label{exchange-series}
\begin{split}
&\mathscr{E}_{ex}=\frac{1}{2}\frac{\ell^2}{h^2}\varepsilon^2\sum_m\Bigl[|\alpha_m|^2\left\langle
f_m'^2+(m^2-1)\frac{f_m^2}{\rho^2}\right\rangle\\
&+|\beta_m|^2\left\langle
g_m'^2+m^2\frac{g_m^2}{\rho^2}\right\rangle\Bigr]\\
&+\sqrt{2}i\frac{\ell^2}{h^2}\varepsilon^2\sum_{m,n}\alpha_m\alpha_n\beta_{m+n}^\star
m\left\langle\frac{f_mf_ng_{m+n}}{\rho^2}\right\rangle+\mathcal{O}(\alpha^4,\beta^4).
\end{split}
\end{equation}
Using that for a weakly excited in-plane vortex state
$\varphi=\chi+\mathfrak{C}\frac{\pi}{2}+\tilde\varphi$,
$m_z=\tilde{m}_z$ the divergence has the form
\begin{equation}\label{eq:div}
\nabla\cdot\vec
m=-\frac{\mathfrak{C}}{\rho}\left[\rho\frac{\partial\tilde\varphi}{\partial\rho}+\tilde
m_z\frac{\partial\tilde
m_z}{\partial\chi}+\tilde\varphi\left(2+\frac{\partial\tilde\varphi}{\partial\chi}\right)\right]+\mathcal{O}(\tilde
m_z^3,\tilde\varphi^3),
\end{equation}
one can perform a similar action to obtain the corresponding
magnetostatic terms:
\begin{equation}\label{eq:en-ms-v-dscr}
\begin{split}
&\mathscr{E}_{ms}^v=\frac{\varepsilon}{2}\sum_m|\beta_m|^2\left\langle\gamma_m(\rho)\gamma_m(\rho')\right\rangle_m^{\mathrm{ms(v)}}\\
&+\frac{i\varepsilon}{\sqrt{2}}\sum_{m,n}m\Biggl[\alpha_n\alpha_{m}\beta_{m+n}^\star\left\langle\gamma_{m+n}(\rho)\frac{f_n(\rho')f_{m}(\rho')}{\rho'}\right\rangle_{m+n}^\mathrm{ms(v)}\\
&+\beta_m\beta_n\beta_{m+n}^\star\left\langle\gamma_{m+n}(\rho)\frac{g_n(\rho')g_{m}(\rho')}{\rho'}\right\rangle_{m+n}^\mathrm{ms(v)}\Biggr]+\mathcal{O}(\alpha^4,\beta^4),
\end{split}
\end{equation}
\begin{equation}\label{eq:en-ms-s-dscr}
\mathscr{E}_{ms}^s=\frac{1}{2\varepsilon}\sum_m|\alpha_m|^2\left\langle
f_m(\rho)f_m(\rho')\right\rangle_m^\mathrm{ms(s)},
\end{equation}
\begin{equation}\label{eq:en-ms-e-dscr}
\begin{split}
&\mathscr{E}_{ms}^e=-\varepsilon\sum_m|\beta_m|^2\left\langle\gamma_m(\rho)g_m(1)\right\rangle_m^{\mathrm{ms(e)}}\\
&-\frac{i\varepsilon}{\sqrt{2}}\sum_{m,n}m\Biggl[\alpha_n\alpha_m\beta_{m+n}^\star\left\langle\frac{f_n(\rho)f_n(\rho)g_{m+n}(1)}{\rho}\right\rangle_{m+n}^\mathrm{ms(e)}\\
&+\beta_n\beta_m\beta_{m+n}^\star\left\langle\frac{g_n(\rho)g_{m}(\rho)g_{m+n}(1)}{\rho}\right\rangle_{m+n}^\mathrm{ms(e)}\Biggr]+\mathcal{O}(\alpha^4,\beta^4).
\end{split}
\end{equation}
The notation
\begin{equation}
\gamma_m(\rho)=\frac{\partial
g_m(\rho)}{\partial\rho}+2\frac{g_m(\rho)}{\rho}
\end{equation}
is used here and three different types of averaging are defined:
\begin{widetext}
\begin{equation}
\begin{split}
&\langle F(\rho,\rho')\rangle_k^\mathrm{ms(v)}\equiv
(-1)^k\int_0^1\mathrm{d}\rho\rho\int_0^1\mathrm{d}\rho'\rho'F(\rho,\rho')\int_0^\infty\frac{e^{-\varepsilon
x}-1+\varepsilon x}{\varepsilon^2 x^2}J_k(\rho
x)J_k(\rho'x)\mathrm{d}x\\
&\langle F(\rho,\rho')\rangle_k^\mathrm{ms(s)}\equiv
(-1)^k\int_0^1\mathrm{d}\rho\rho\int_0^1\mathrm{d}\rho'\rho'F(\rho,\rho')\int_0^\infty(1-e^{-\varepsilon
x})J_k(\rho x)J_k(\rho'x)\mathrm{d}x\\
&\langle F(\rho)\rangle_k^\mathrm{ms(e)}\equiv
(-1)^k\int_0^1\mathrm{d}\rho\rho
F(\rho)\int_0^\infty\frac{e^{-\varepsilon x}-1+\varepsilon
x}{\varepsilon^2 x^2}J_k(\rho x)J_k(x)\mathrm{d}x.
\end{split}
\end{equation}
Comparing \eqref{eq:en-ms-v-dscr}--\eqref{eq:en-ms-e-dscr} with
\eqref{eq:en-osc} and \eqref{eq:en-int} one can conclude that the
eigenfrequencies read
\begin{equation}
\begin{split}
\varOmega_m=&\frac{\ell^2}{h^2}\varepsilon^2\left\langle
f_m'^2+(m^2-1)\frac{f_m^2}{\rho^2}\right\rangle
+\frac{1}{\varepsilon}\left\langle
f_m(\rho)f_m(\rho')\right\rangle_m^\mathrm{ms(s)}\\
=&\frac{\ell^2}{h^2}\varepsilon^2\left\langle
g_m'^2+m^2\frac{g_m^2}{\rho^2}\right\rangle+\varepsilon\left\langle\gamma_m(\rho)\gamma_m(\rho')\right\rangle_m^{\mathrm{ms(v)}}-2\varepsilon\left\langle\gamma_m(\rho)g_m(1)\right\rangle_m^{\mathrm{ms(e)}},
\end{split}
\end{equation}
and the nonlinearity coefficients have form
\begin{equation}
\begin{split}
k_{m,n}^\alpha=&\sqrt{2}\frac{\ell^2}{h^2}\varepsilon^2\left\langle\frac{f_mf_ng_{m+n}}{\rho^2}\right\rangle+\frac{\varepsilon}{\sqrt{2}}\left[\left\langle\gamma_{m+n}(\rho)\frac{f_n(\rho')f_{m}(\rho')}{\rho'}\right\rangle_{m+n}^\mathrm{ms(v)}-\left\langle\frac{f_n(\rho)f_n(\rho)g_{m+n}(1)}{\rho}\right\rangle_{m+n}^\mathrm{ms(e)}\right],\\
k_{m,n}^\beta=&\frac{\varepsilon}{\sqrt{2}}\left[\left\langle\gamma_{m+n}(\rho)\frac{g_n(\rho')g_{m}(\rho')}{\rho'}\right\rangle_{m+n}^\mathrm{ms(v)}-\left\langle\frac{g_n(\rho)g_n(\rho)g_{m+n}(1)}{\rho}\right\rangle_{m+n}^\mathrm{ms(e)}\right].
\end{split}
\end{equation}
\end{widetext}
It is interesting to emphasize the following properties of the
energy expansions: (i) The linear part of the magnetostatic energy
proportional to $|\beta|^2$ is produced by the volume charges, while
the part proportional to $|\alpha|^2$ is produced by surface
charges; (ii) The nonlinear terms of the magnetostatic energy appear
due to the volume charges only; (iii) The exchange produces
nonlinear terms of the form $\alpha\alpha\beta$. The same is true
for magnetostatics, but additional terms in the form
$\beta\beta\beta$ also appear.
%


\end{document}